\documentclass[a4paper,11pt]{article}
\pdfoutput=1
\usepackage{jcappub}
\usepackage{slashed}
\usepackage{bm} 
\usepackage{graphicx,epsf,epstopdf}
\usepackage{latexsym,amssymb,amsmath,float,url,mathrsfs}
\usepackage{array}
\usepackage{soul}
\usepackage{natbib}
\usepackage{hyperref}
\usepackage{slashed}
\usepackage{subfigure}
\usepackage[dvipsnames]{xcolor}

\hypersetup{
     colorlinks   = true,
     citecolor    = violet,
     urlcolor     = violet,
     linkcolor    = violet
}

\input{epsf}

\makeatletter
\newcommand{\pushright}[1]{\ifmeasuring@#1\else\omit\hfill$\displaystyle#1$\fi\ignorespaces}
\newcommand{\pushleft}[1]{\ifmeasuring@#1\else\omit$\displaystyle#1$\hfill\fi\ignorespaces}
\makeatother

\begin{document}

\title{\boldmath Dark Forces in the Sky: Signals from $Z'$ and the Dark Higgs }

\author{Nicole F.\ Bell,}
\author{Yi Cai and}
\author{Rebecca K.\ Leane}
\affiliation{ARC Centre of Excellence for Particle Physics at the Terascale \\
School of Physics, The University of Melbourne, Victoria 3010, Australia}

\emailAdd{\tt n.bell@unimelb.edu.au}
\emailAdd{\tt yi.cai@unimelb.edu.au}
\emailAdd{\tt rleane@physics.unimelb.edu.au}

\date{\today}

\abstract{We consider the indirect detection signals for a
  self-consistent hidden $U(1)$ model containing a Majorana dark
  matter candidate, $\chi$, a dark gauge boson, $Z'$, and a dark
  Higgs, $s$.  Compared with a model containing only a dark matter
  candidate and $Z'$ mediator, the addition of the scalar provides a
  mass generation mechanism for the dark sector particles and is
  required in order to avoid unitarity violation at high energies.  We
  find that the inclusion of the two mediators opens up a new two-body
  $s$-wave annihilation channel, $\chi \chi\rightarrow sZ'$.  This new
  process, which is missed in the usual single-mediator simplified
  model approach, can be the dominant annihilation channel.  This
  provides rich phenomenology for indirect detection searches, allows
  indirect searches to explore regions of parameter space not
  accessible with other commonly considered $s$-wave annihilation
  processes, and enables both the $Z'$ and scalar couplings to be
  probed.  We examine the phenomenology of the sector with a focus on
  this new process, and determine the limits on the model parameter
  space from Fermi data on dwarf spheriodal galaxies and other
  relevant experiments.}

\maketitle


\section{Introduction}

While dark matter (DM) is thought to be the dominant form of matter in
the universe, its fundamental nature remains unknown.  Of the many
possible types of DM candidates, a particularly well motivated choice
are Weakly Interacting Massive Particles
(WIMPs)~\cite{Bergstrom:2000pn,Bertone:2004pz}.  This class of DM
contains an abundance of models. In order to discover which of the
many models may be the correct description, it is necessary to make
contact between these theories and experiments.  To efficiently test
many of these models, it is desirable to investigate the properties of
DM in a model independent manner wherever possible. This is reasonably
achieved within the simplified model
framework~\cite{Abdallah:2014hon,Buckley:2014fba,Alves:2011wf,Alwall:2008ag,Abdallah:2015ter,Abercrombie:2015wmb,DeSimone:2016fbz,Jacques:2016dqz},
where only the lightest particles in the theory are retained, and they
can be generically explored via phenomenologically distinct couplings
and mediator choices. Specifically, the three benchmark simplified
models for DM and Standard Model (SM) interactions are a spin-1
mediated $s$-channel interaction, a spin-0 mediated $s$-channel
interaction, and a spin-0 mediated $t$-channel
interaction~\cite{Abercrombie:2015wmb}.

However, due to their simplified nature and reduced number of
parameters, these benchmark models are not intrinsically capable of
capturing the full phenomenology of many realistic UV complete
theories.  Perhaps worse is that the separate consideration of these
benchmarks can lead to physical problems and inconsistencies.  For
instance, the consequences of gauge invariance and unitarity violation
have recently been discussed
in~\cite{Shoemaker:2011vi,Fox:2012ee,Busoni:2013lha,Buchmueller:2013dya,Busoni:2014sya,Endo:2014mja,Busoni:2014haa,Hedri:2014mua,Bell:2015sza,Baek:2015lna,Kahlhoefer:2015bea,Bell:2015rdw,Haisch:2016usn,Englert:2016joy,Bell:2016obu}.

\begin{figure}[t]
\centering
\subfigure{\includegraphics[width=0.28\columnwidth]{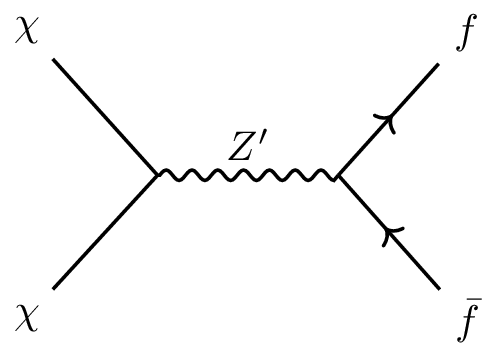}}
\hspace{1mm}
\subfigure{\includegraphics[width=0.32\columnwidth]{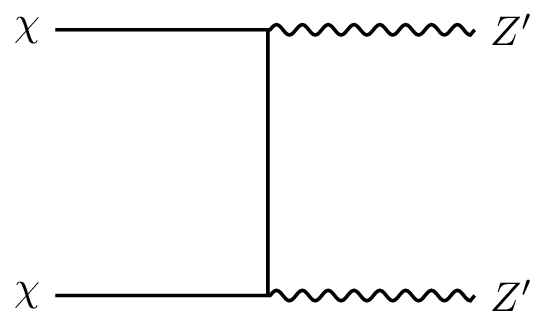}}
\caption{Spin-1 simplified model annihilation processes. Left: This
  process has an $s$-wave component only if the mediator has
  axial-vector couplings to SM fermions, $f$. However, the
  non-observation of a direct detection or LHC signal makes it
  difficult to obtain a thermal relic scale cross section from this
  diagram. Right: This process is $s$-wave for all field or coupling
  types and, as it can avoid LHC and direct detection bounds in the
  hidden on-shell mediator scenario, is often considered in the
  indirect detection context.}
\label{fig:simpmodelvec}
\end{figure}

\begin{figure}[t]
\centering
\subfigure{\includegraphics[width=0.28\columnwidth]{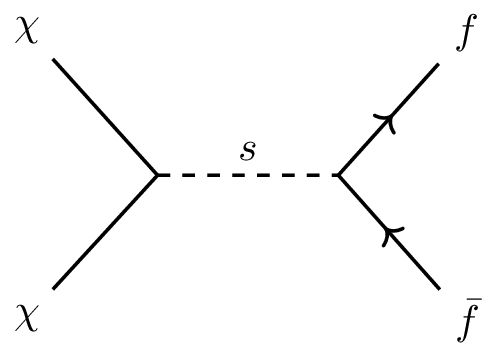}}
\hspace{1mm}
\subfigure{\includegraphics[width=0.32\columnwidth]{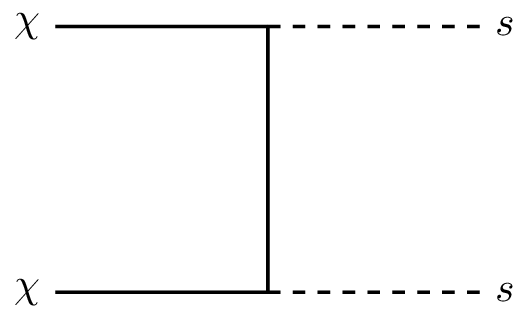}}
\hspace{1mm}
\subfigure{\includegraphics[width=0.32\columnwidth]{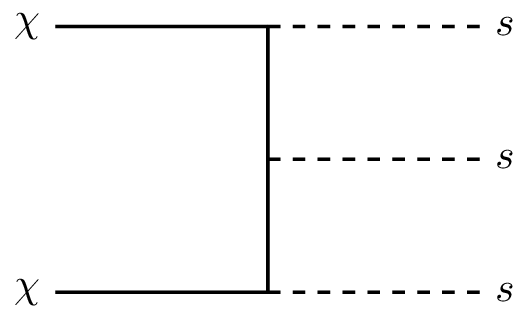}}
\caption{Spin-0 simplified model annihilation processes. Left: This
  process has an $s$-wave component if the spin-0 field is a
  pseudoscalar. However, the non-observation of a direct detection or
  LHC signal makes it difficult to achieve a thermal relic density
  with this process. Middle: This process is $p$-wave for all field or
  coupling types. Right: This process has an $s$-wave contribution if
  the spin-0 field is a pseudoscalar, but it is three-body phase space
  suppressed. There is no $s$-wave process for fermionic DM
  annihilation to a spin-0 field with scalar couplings.}
\label{fig:ss}
\end{figure}

These issues motivate a scenario in which the vector and the scalar
mediators appear together within the same theory\footnote{Some recent
  work on multi-mediator models can be found in
  Refs. \cite{Cline:2015qha,Choudhury:2015lha,Ghorbani:2015baa,Duerr:2016tmh}.}.
Specifically, a simplified model with a spin-1 mediator and
axial-vector couplings to fermions will lead to unitarity violation at
high energies unless some additional new physics, such a scalar degree
of freedom, is introduced to the simplified model
setup~\cite{Kahlhoefer:2015bea}.  This scalar is exceedingly well
motivated if it is also taken to provide a mass generation mechanism
for the dark sector, as the ``dark Higgs''.
The purpose of this paper is to explore the indirect detection signals
for a gauge invariant model where the dark sector consists of a
fermionic DM candidate, a spin-1 mediator, and a dark Higgs field.
In doing so, we shall encounter important phenomenology that
cannot be captured by a single-mediator model.

In the indirect detection context, simplified models have commonly
been used to investigate annihilation processes and place limits on
the dark matter parameter space.  Only annihilations which proceed via
an $s$-wave process contribute substantially to DM signals in the
universe today, as $p$-wave contributions are highly suppressed by a
velocity squared factor, $v_\chi^2\approx10^{-6}$.  Within the
simplified model framework, spin-1 mediators provide two possible
two-body $s$-wave annihilation processes for fermionic dark matter, as
shown in Fig.~(\ref{fig:simpmodelvec}). (i)
$\chi {\chi}\rightarrow f {f}$ has an $s$-wave
component provided the mediator has axial-vector couplings to SM
fermions, $f$ while (ii) $\chi {\chi}\rightarrow Z'Z'$ has an
$s$-wave component for any (vector or axial-vector) coupling of the
$Z'$ to $\chi$.  The latter process, with the $Z'$ pair produced
on-shell, is commonly studied in the indirect detection context; it is
capable of producing large annihilation signals while avoiding strong
constraints imposed by collider and direct detection searches.

For spin-0 mediators, $\chi {\chi}\rightarrow f {f}$
is $s$-wave if the mediator is a pseudoscalar, but the couplings to SM
fermions are strongly constrained, such that a thermal relic cross
section is not easily obtained, nor a large indirect detection
signal. The remaining 2-body annihilation processes for spin-0
mediators are all $p$-wave, meaning that to obtain a non-negligble
indirect detection signal with non-excluded parameters, one needs to
resort to the case where three spin-1 fields, $s$, are
produced\footnote{A two-body $s$-wave process is possible for
  combinations of multiple distinct scalars \cite{Berlin:2014pya,
    Abdullah:2014lla, Chao:2016cea}, but this extends beyond the
  simplified model framework and requires more detailed model
  building.} as $\chi {\chi}\rightarrow sss$. While this is
an $s$-wave process provided that the mediator is a pseudoscalar, it
suffers from three-body phase space suppression
\cite{Abdullah:2014lla}.  These processes are shown in
Fig.~(\ref{fig:ss}).

In this paper, we will show that once the dark Higgs is added to the
dark sector, the indirect detection phenomenology considered
previously was incomplete. Of particular interest will be the new
$s$-wave annihilation process,
\begin{equation}
\chi {\chi}\rightarrow s Z'.
\end{equation} 
This is always an $s$-wave process, irrespective of whether the
DM-$Z'$ coupling is vector or axial-vector, and irrespective of
whether $s$ is a scalar or pseudoscalar.  This process allows for new,
rich phenomenology.  It allows the spin-0 particle to play an
important role in indirect detection, which is not possible in models
with only a spin-0 mediator due to the velocity or phase space
suppressions of the annihilation diagrams in the pseudoscalar mediator
case, and the complete absence of any $s$-wave annihilation processes
in the scalar mediator case. Importantly, although both the
$\chi\chi\rightarrow sZ'$ and $\chi\chi\rightarrow Z'Z'$ annihilation
channels have an s-wave component, the $sZ'$ channel tends to dominate
when it is kinematically accessible.  Neglecting this important
annihilation process would lead to dramatically different results.

Hidden sector
models~\cite{Pospelov:2007mp,Pospelov:2008jd,Pospelov:2008zw,Feng:2008mu,Feng:2008ya,
  Rothstein:2009pm,Mardon:2009gw,Mardon:2009rc,
  Meade:2009rb,Cheung:2010gj, Davoudiasl:2013jma,
  Berlin:2014pya,Liu:2014cma, Hardy:2014dea,
  Boehm:2014bia,McDermott:2014rqa,Chacko:2015noa, Elor:2015tva,
  Elor:2015bho, Abdullah:2014lla, Martin:2014sxa, Ko:2015ioa,
  Ko:2014gha, Kim:2016csm, Hooper:2012cw,
  Berlin:2015wwa,Cline:2014dwa,Morrissey:2014yma} are a specific
realization of simplified models, commonly adopted in the indirect
detection scenario because their small direct couplings to the SM
ameliorate the tension between strong constraints from collider and
direct detection experiments, and the goal of a sizeable indirect
detection signal.  If the DM annihilates to on-shell mediators (rather
than directly to SM particles via off-shell mediators) the smallness
of the dark-SM couplings are irrelevant for indirect detection,
provided of course that the dark-sector mediators eventually decay to
visible sector particles with lifetime shorter than the age of the
galaxy. The signal size for indirect detection is instead set by the
size of the dark sector couplings, which can often be taken to be
quite large.

In this paper, we will investigate the phenomenology of these indirect
detection signals for a self-consistent hidden $U(1)$ sector, with a
focus on the impact of this new $\chi \chi \rightarrow sZ'$
annihilation channel. In Section \ref{sec:model}, we will describe the
model in detail. We will then list all the annihilation processes of
interest in this model, along with the relevant cross sections and decay
widths, in Section \ref{sec:signals}. In Section \ref{sec:spectra}, we
will simulate the consequent $\gamma$-ray spectra, which we will use
in Section \ref{sec:dwarfs} to calculate the limits on the cross
section and parameter space from Fermi-LAT data on dwarf spheriodal
galaxies, the most dark matter dense objects in our sky, as well as
AMS-02. Finally we will consider relevant limits from unitarity and
other experiments in Section \ref{sec:other}, and summarize in Section
\ref{sec:summary}.


\section{Model Setup}
\label{sec:model}

The gauge symmetry group for our model is $SU(3)_c\otimes
SU(2)_W\otimes U(1)_Y\otimes U(1)_\chi$, such that the covariant
derivative is $D_\mu=D_\mu^{SM}+i Q' g_{\chi}Z'_\mu$ with $Q'$ being
the dark $U(1)_\chi$ charge of the relevant fields.  We introduce new
fields: a Majorana fermion DM candidate $\chi$, a spin-1 dark gauge
boson $Z'$, and the dark Higgs field $S$. 
We have chosen $\chi$ to be Majorana, as a well-motivated example that
must involve axial-vector couplings to the $Z'$, given that vector
couplings of Majorana particles vanish.  The significance of
axial-vector couplings is that perturbative unitarity would be
violated at high energy in the absence of
$S$~\cite{Kahlhoefer:2015bea}. The dark Higgs is mandatory in this
set-up.

The vacuum expectation value (vev) of the dark Higgs field provides a
mass generation mechanism for the dark sector fields $Z'$ and
$\chi$. For the $\chi$-$S$ Yukawa terms to respect the $U(1)_\chi$
gauge symmetry, the charge assignments\footnote{In order to cancel
  anomalies, additional fermions with $U(1)_\chi$ charge will be
  required. However, these states can be made sufficiently heavy that
  they do not affect by the dark sector phenonenology discussed
  here. For example, anomaly cancellation can be achieved by
  introducing an additional Majorana fermion, with $U(1)_\chi$ charge
  equal in magnitude but of opposite sign to that of $\chi$.  It is
  sufficient to consider only the lighter of the two fermions as the
  DM candidate, with the heavier making a subdominant contribution to
  the relic density~\cite{Cline:2014dwa}.}  can be chosen, without
loss of generality, to be $Q'(S)=1$ and $Q'(\chi)=-\frac{1}{2}$.  The
dark Higgs can mix with the SM Higgs $H$ through mass mixing, with
strength parameterized by $\lambda_{hs}$, while the $U(1)_\chi$ field
strength tensor $Z'_{\mu\nu}$ kinetically mixes with the SM
hypercharge field strength $B_{\mu\nu}$ controlled by the kinematic
mixing parameter $\epsilon$.  Explicitly, before electroweak and dark
symmetry breaking, the Lagrangian is written as
\begin{eqnarray}
 \mathcal{L}&=&\mathcal{L}_{SM}+\frac{i}{2}\overline{\chi}\slashed{\partial}\chi - \frac{1}{4} g_\chi Z'^\mu\overline{\chi}\gamma_5\gamma_\mu\chi
- \frac{1}{2} y_\chi \overline{\chi} \left( P_L S + P_R S^* \right) \chi \; 
 -\frac{\sin\epsilon}{2} Z'^{\mu\nu}B_{\mu\nu}\\
 &+& \left[ (\partial^\mu + ig_\chi Z'^\mu ) S \right]^\dagger \left[ (\partial_\mu + ig_\chi Z'_\mu ) S \right] - \mu_s^2 S^\dagger S - \lambda_s (S^\dagger S)^2
 - \lambda_{hs}(S^\dagger S)(H^\dagger H). \nonumber
\end{eqnarray}
After symmetry breaking and mixing the terms of interest are
\begin{eqnarray}
 \mathcal{L}&\supset&\frac{1}{2}m_{Z'}^2  Z'^\mu Z'_\mu -\frac{1}{2}m_{s}^2 s^2 - \frac{1}{2} m_{\chi}\overline{\chi}\chi -\frac{1}{4} g_\chi Z'^\mu\overline{\chi}\gamma_5\gamma_\mu \chi - \frac{y_\chi}{2\sqrt{2}}s\overline{\chi}\chi \nonumber \\
 &+& g_\chi^2 w Z'^\mu Z'_\mu s - \lambda_s w s^3 - 2\lambda_{hs} (hvs^2+swh^2)+g_f \sum_f Z'^\mu \overline{f}\Gamma_\mu f, 
\label{eq:brokenlgn}
 \end{eqnarray}
where the component fields of $S$ and $H$ are defined in the broken
phase as $S\equiv \frac{1}{\sqrt{2}}(w + s + i a)$ and $H=\left\{ G^+,
\frac{1}{\sqrt{2}}(v + h +i G^0)\right\}$ with $G^+$, $G^0$ and $a$
being the Goldstone bosons of $W$, $Z$ and $Z^\prime$ respectively,
while $s$ and $h$ are real scalars.  In the limit that the mixing
parameter $\lambda_{hs}$ is small, the vev of
the dark Higgs satisfies $w^2=-\mu_s^2/\lambda_s$. After symmetry
breaking, the masses are
\begin{subequations}
\begin{equation}
   m_{Z'}=g_\chi w,
 \end{equation}
 \begin{equation}
   m_{\chi}=\frac{1}{\sqrt{2}}y_\chi w,
 \end{equation}
  \begin{equation}
   m_{s}^2 \simeq -\mu_s^2,
 \end{equation}
  \begin{equation}
   m_{h} \simeq -\mu_h^2.
\end{equation}
\label{eq:masses}
\end{subequations}
For all couplings to remain perturbative, only certain combinations of
the dark gauge coupling and dark sector masses are allowed. From the
above equations, the relation between the dark yukawa coupling
$y_\chi$ and the $U(1)_\chi$ gauge coupling $g_\chi$ is
\begin{equation}
   \frac{y_{\chi}}{g_\chi}=\frac{\sqrt{2}\;m_\chi}{m_{Z'}}.
   \label{eq:yukawa}
 \end{equation}

The final term of Eq.~(\ref{eq:brokenlgn}) describes the coupling of
$Z'$ to the SM fermions; its structure is dictated by the kinetic
mixing, and the explicit form can be found, for example, in
Ref.~\cite{Agashe:2014kda}.  As $Z'$ decays to the SM through the
hypercharge portal, the $Z'$ couples to the same SM fields as the SM
$Z$, and no flavor specific tuning is permitted.  This enforces strong
di-lepton resonance bounds and EWPT limits on $Z$-$Z'$
mixing. Regardless, the small values of $\epsilon$ we consider allow
these bounds to be easily satisfied.

Within this model, there are two possible routes for dark sector
interactions with the visible sector: the Higgs portal controlled by
parameter $\lambda_{hs}$, or the hypercharge portal controlled by
parameter $\epsilon$. To demonstrate the new phenomenology of the
combination of both the $Z'$ and dark Higgs in this model, we will
take small values of these parameters consistent with the hidden model
setup, and assume both $s$ and $Z'$ decay on-shell to SM fermions.  As
the Higgs couples to fields proportional to their masses, the dark
Higgs decays predominantly to $b$-quarks in the mass range we
consider, although we will fully simulate all final states.  The dark
Higgs may also decay into two $Z'$ which then may decay into SM
fermions, however for simplicity when setting limits we will focus on
the region of parameter space where this is not kinematically allowed.

We emphasize that this is the most general scenario involving the
interaction of a Majorana fermion with a $Z'$ gauge boson.  Given that
vector currents vanish for Majorana fermions, leaving only
axial-vector interactions, the inclusion of the dark Higgs is
unavoidable in order to provide a mass for the $Z'$ within a gauge
invariant model that respects perturbative unitarity.  Furthermore, it
is not possible to include a Majorana mass term for the $\chi$ without
breaking the $U(1)_\chi$ symmetry.  The case of Dirac dark matter with
vector couplings to a $Z'$ would be very different.  In that case, the
$Z'$ may obtain mass via the Stueckelberg mechanism, and a bare mass
term for $\chi$ is possible, leaving no need for a dark Higgs.

\section{Dark Matter Annihilation Processes for Indirect Detection}
\label{sec:signals}

In this section we will calculate the annihilation cross sections and
branching fractions relevant for indirect detection.

\subsection{Annihilation Cross Sections}

\begin{figure}[t]
\centering
\subfigure{\includegraphics[width=0.28\columnwidth]{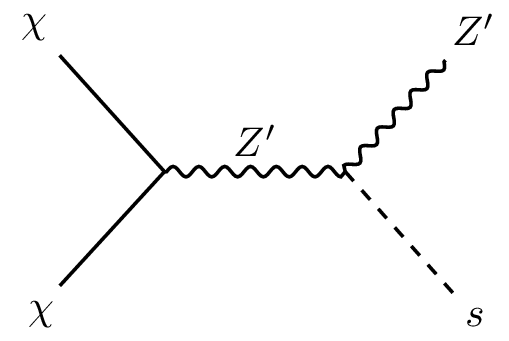}}
\hspace{1mm}
\subfigure{\includegraphics[width=0.32\columnwidth]{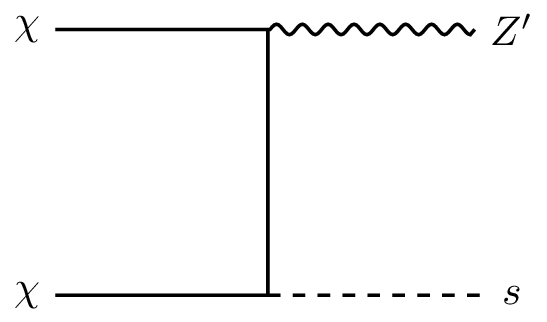}}
\hspace{1mm}
\subfigure{\includegraphics[width=0.32\columnwidth]{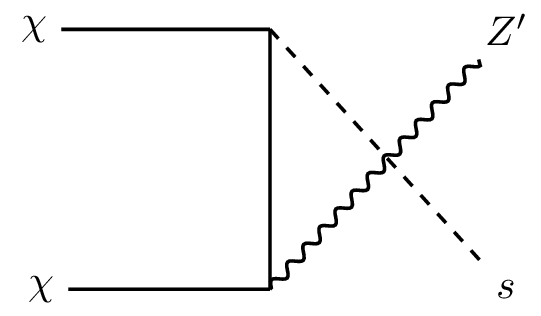}}
\caption{Annihilation diagrams for the $s$-wave processes
  $\chi {\chi}\rightarrow sZ'$. The scalar and the $Z'$ then
  can decay to SM fermion final states. For some regions of parameter
  space this is the only kinematically allowed process, while in
  others it can have a cross section larger than the process in
  Fig.~(\ref{fig:zpzp}). This process can be achieved by considering
  the simplified model benchmarks together.}
\label{fig:zps}
\end{figure}

The novel process for DM annihilation in the universe today is
$\chi {\chi}\rightarrow s Z'$, which is shown in
Fig.~(\ref{fig:zps}).  This process has not been considered in
previous work, despite being a consequence of a self-consistent $Z'$
model with axial-vector couplings.  The cross section for
$\chi {\chi}\rightarrow s Z'$ is $s$-wave for both scalar and
pseudoscalar interactions, and vector or axial-vector $Z'$-DM
couplings.  For Majorana DM and a real scalar the annihilation cross
section is given by
\begin{equation}
\langle\sigma v\rangle_{\chi {\chi}\rightarrow s Z'} = 
\frac{g_\chi^4 \left(m_s^4-2 m_s^2 \left(m_{Z'}^2+4 m_{\chi}^2\right)
+\left(m_{Z'}^2-4 m_{\chi }^2\right)^2\right)^{3/2}}{1024 \pi  m_{\chi}^4 m_{Z'}^4},
 \label{eq:xseczpsA}
\end{equation}
where Eq.~(\ref{eq:yukawa}) has been used to replace $y_\chi$.
Here, only the $s$-channel diagram of Fig.~(\ref{fig:zps}) contributes
an $s$-wave component.

\begin{figure}[t]
\centering
\subfigure{\includegraphics[width=0.28\columnwidth]{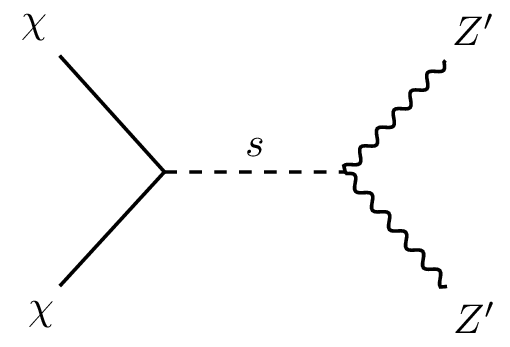}}
\hspace{1mm}
\subfigure{\includegraphics[width=0.32\columnwidth]{tchan_zpzp.png}}
\hspace{1mm}
\subfigure{\includegraphics[width=0.32\columnwidth]{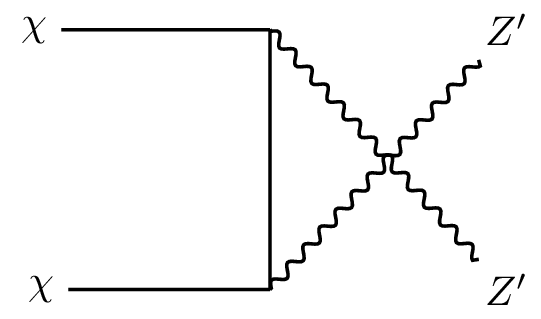}}
\caption{Annihilation diagrams for the $s$-wave processes
  $\chi {\chi}\rightarrow Z'Z'$. The $Z'$ then can decay into
  SM fermion final states. In the spin-1 mediator simplified model
  benchmark, only the $t$-channel and $u$-channel diagrams appear,
  leading to unitarity issues at high energies for axial couplings. In
  our gauge invariant model, the $s$-channel diagram restores
  perturbative unitarity. Consideration of only
  $\chi {\chi}\rightarrow Z'Z'$, without the accompanying
  $\chi {\chi}\rightarrow sZ'$ process of Fig.~(\ref{fig:zps})
  will lead to inaccurate conclusions.}
\label{fig:zpzp}
\end{figure}

The other dominant $s$-wave process in this model is
$\chi {\chi}\rightarrow Z' Z'$, which is shown in
Fig.~(\ref{fig:zpzp}). For Majorana DM, the $s$-wave contribution to
its cross section is given by\footnote{The factor of 16 difference
  between our cross section and that given in other papers is due
  to the $(Q'_\chi)^4=(1/2)^4$ charge contribution to the coefficient.}
\begin{equation}
\langle\sigma v\rangle_{\chi {\chi}\rightarrow Z' Z'} = \frac{g_\chi^4\left(1-\frac{m_{Z'}^2}{m_\chi^2}\right)^{3/2}}{256\pi m_\chi^2\left(1-\frac{m_{Z'}^2}{2m_\chi^2}\right)^2} \; ,
 \label{eq:xseczpzp}
\end{equation}
where the $s$-wave contributions only come from the $t$ and $u$ channel diagrams, making the indirect signal for the $Z'Z'$ process the same as that found in the spin-1
simplified model benchmark.

Previously, annihilation of fermionic dark matter to spin-0 mediators
featured an s-wave component only for the three-body phase-space
suppressed process in Fig.~(\ref{fig:ss}), and only for pseudoscalars.
For a simplified model with a scalar mediator, there is no s-wave
annihilation process at all.  We make the important observation that
annihilation of fermionic dark matter to a spin-0 plus spin-1 final
state will always be $s$-wave, for both scalars and pseudoscalars.
This allows indirect detection to have comparable sensitivity for
spin-0 and spin-1 mediators, in models where the two mediators are
both present.  This is realized naturally in the very simple gauge
invariant model we have presented in this paper.

\begin{figure}[t]
     \begin{center}
        \subfigure{
            \includegraphics[width=0.4799\columnwidth]{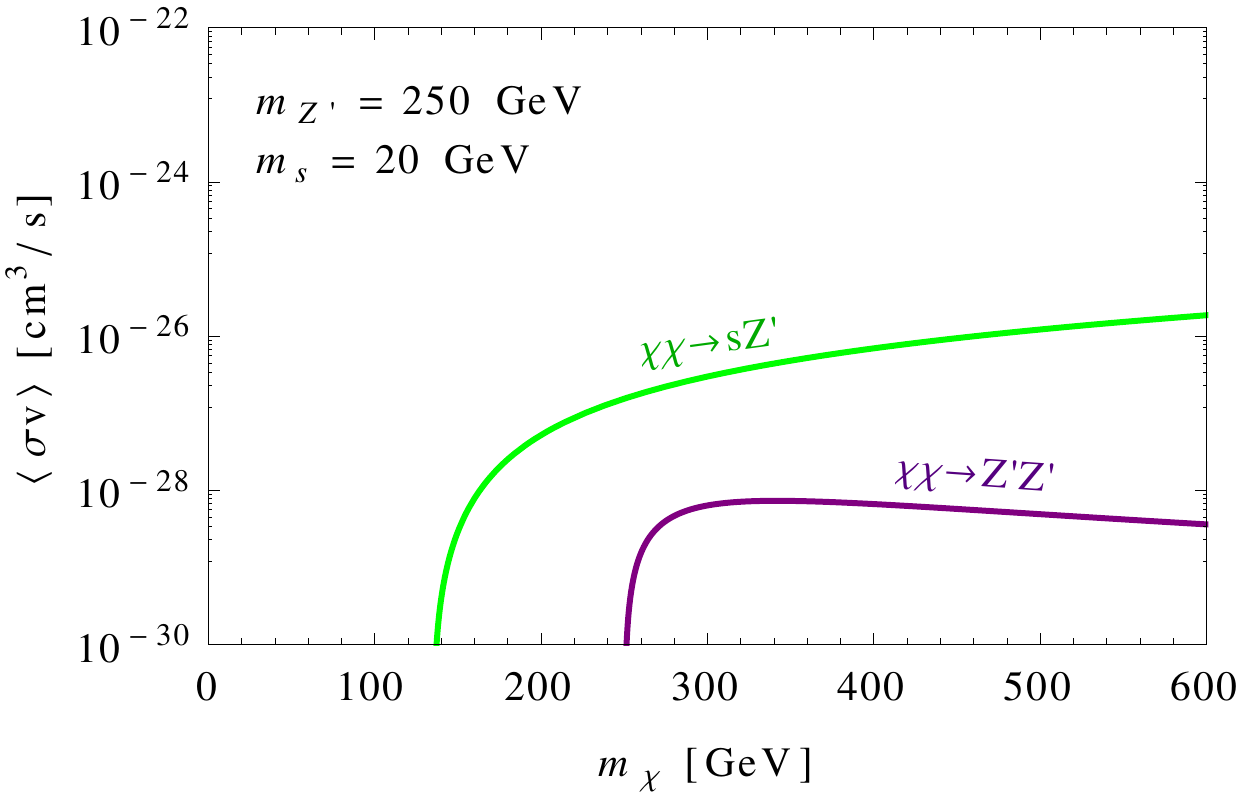}}
        \subfigure{
           \includegraphics[width=0.4799\columnwidth]{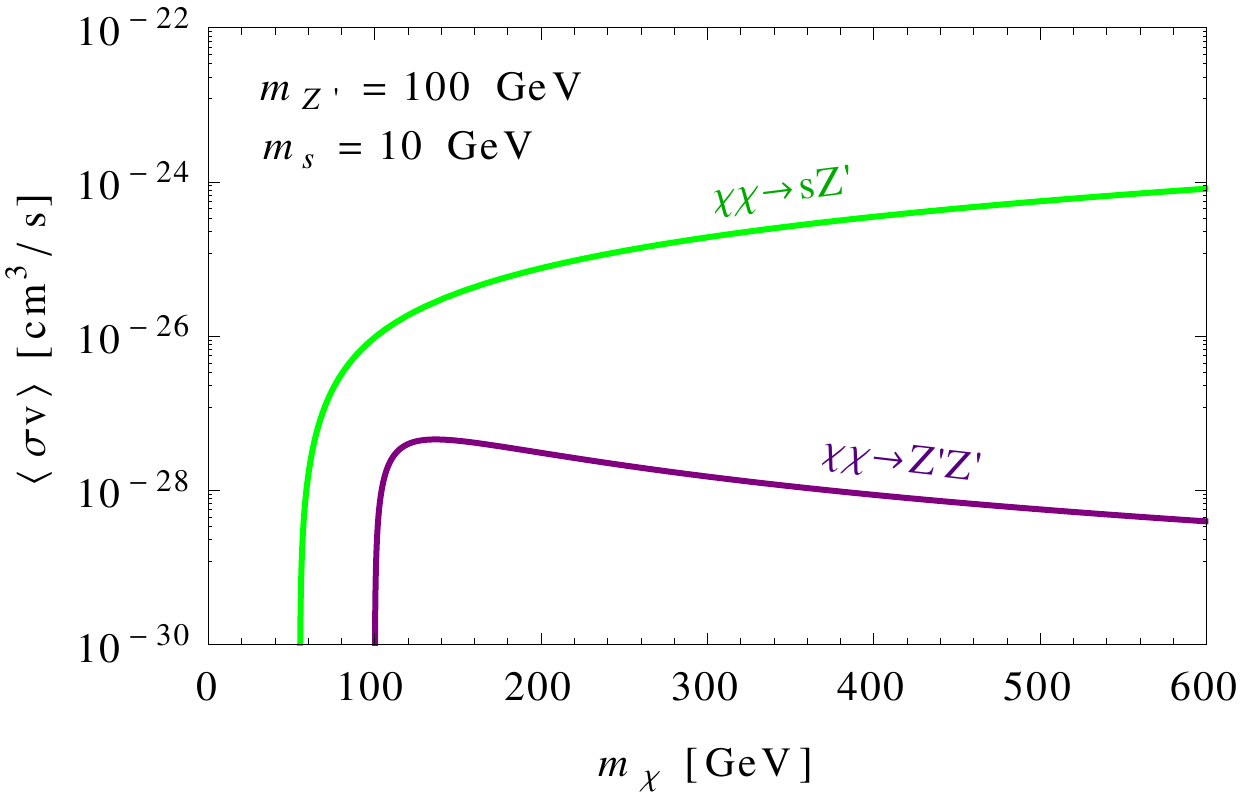}}
        \\ 
        \subfigure{
            \includegraphics[width=0.4799\columnwidth]{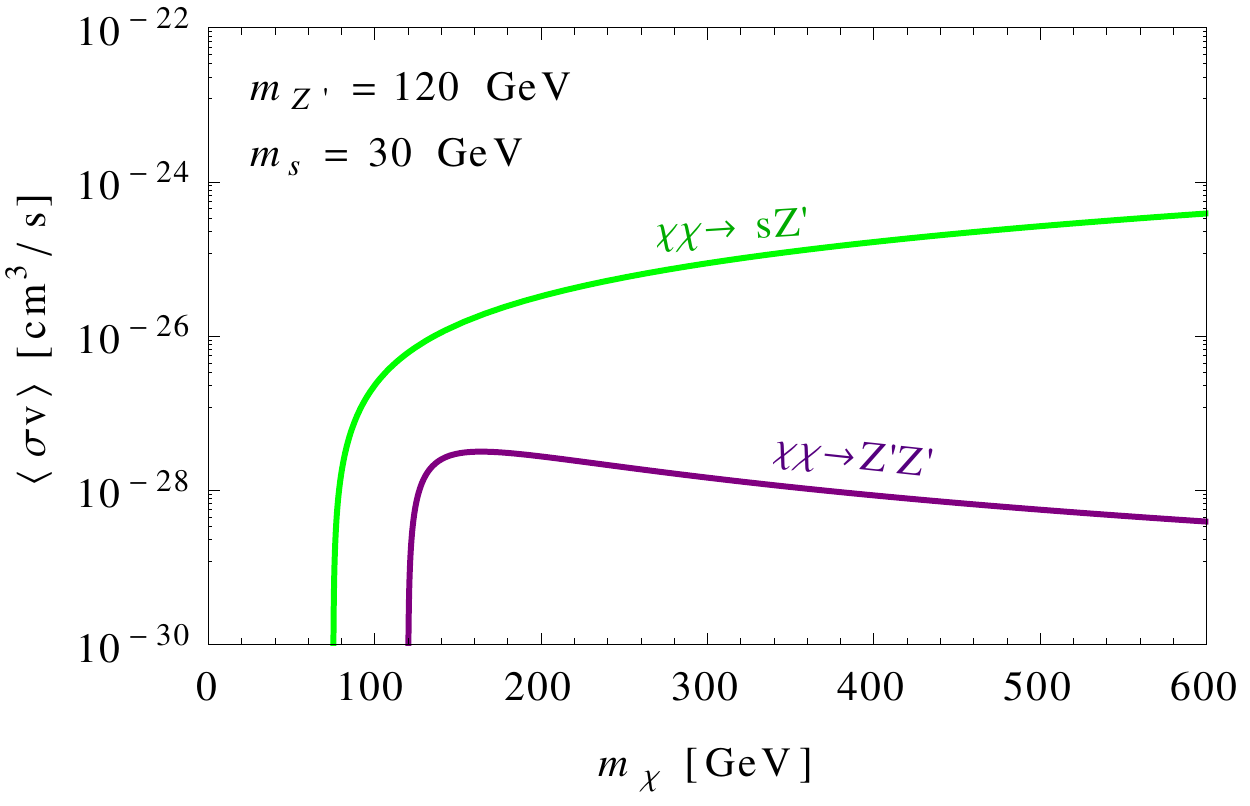}}
        \subfigure{
            \includegraphics[width=0.4799\columnwidth]{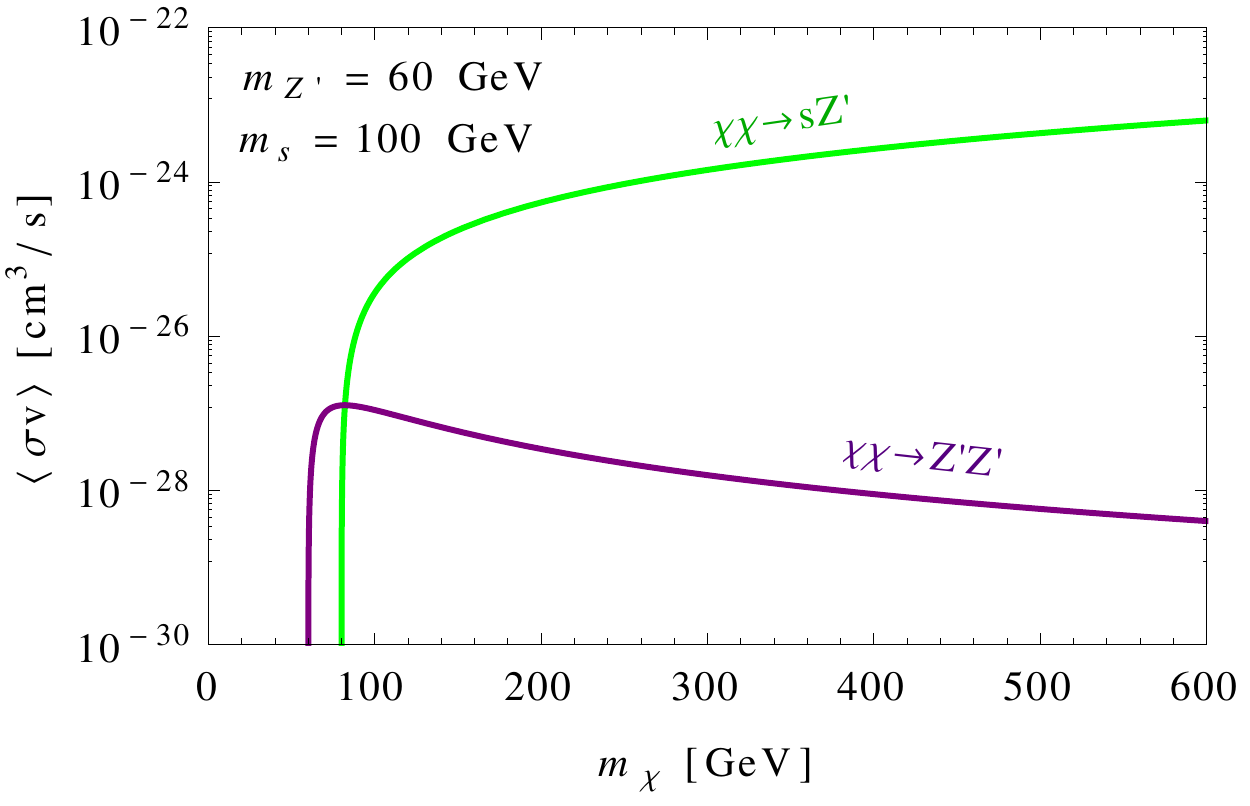}}
        \\
         \subfigure{
            \includegraphics[width=0.479\columnwidth]{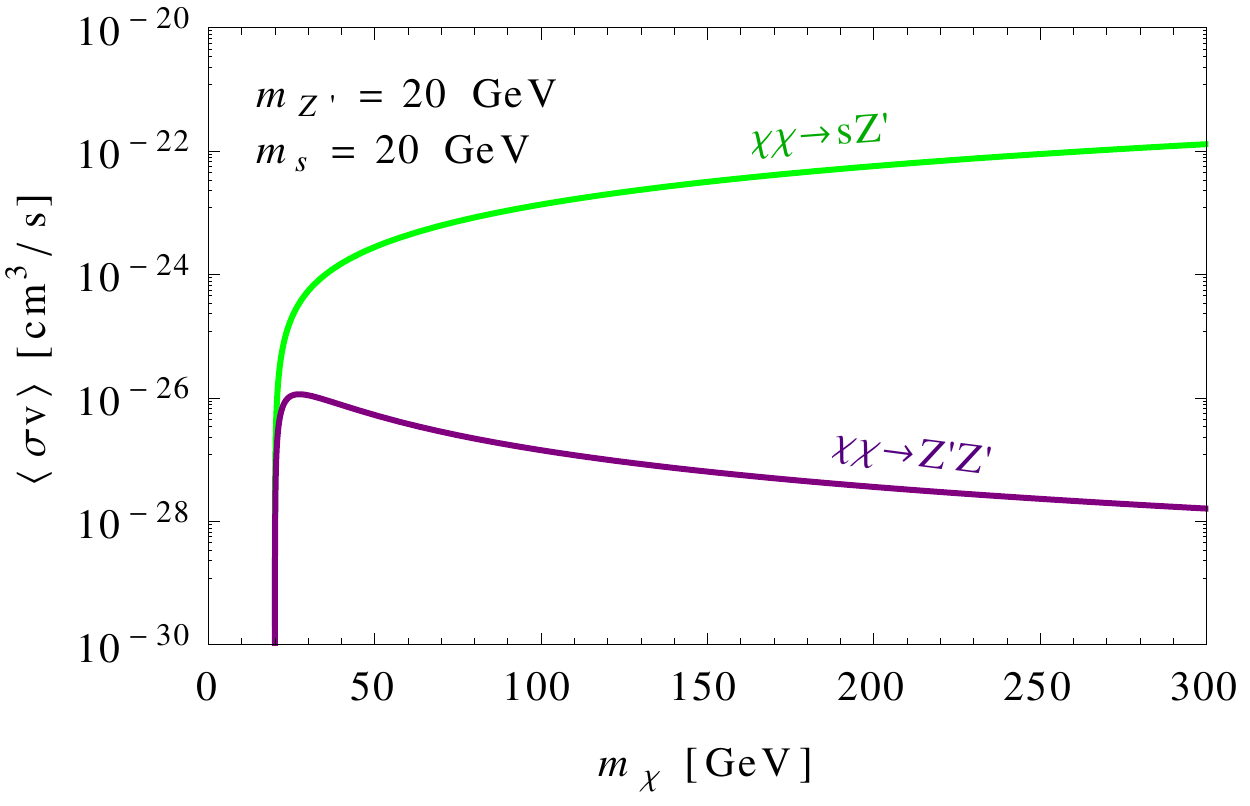}}
         \subfigure{
            \includegraphics[width=0.479\columnwidth]{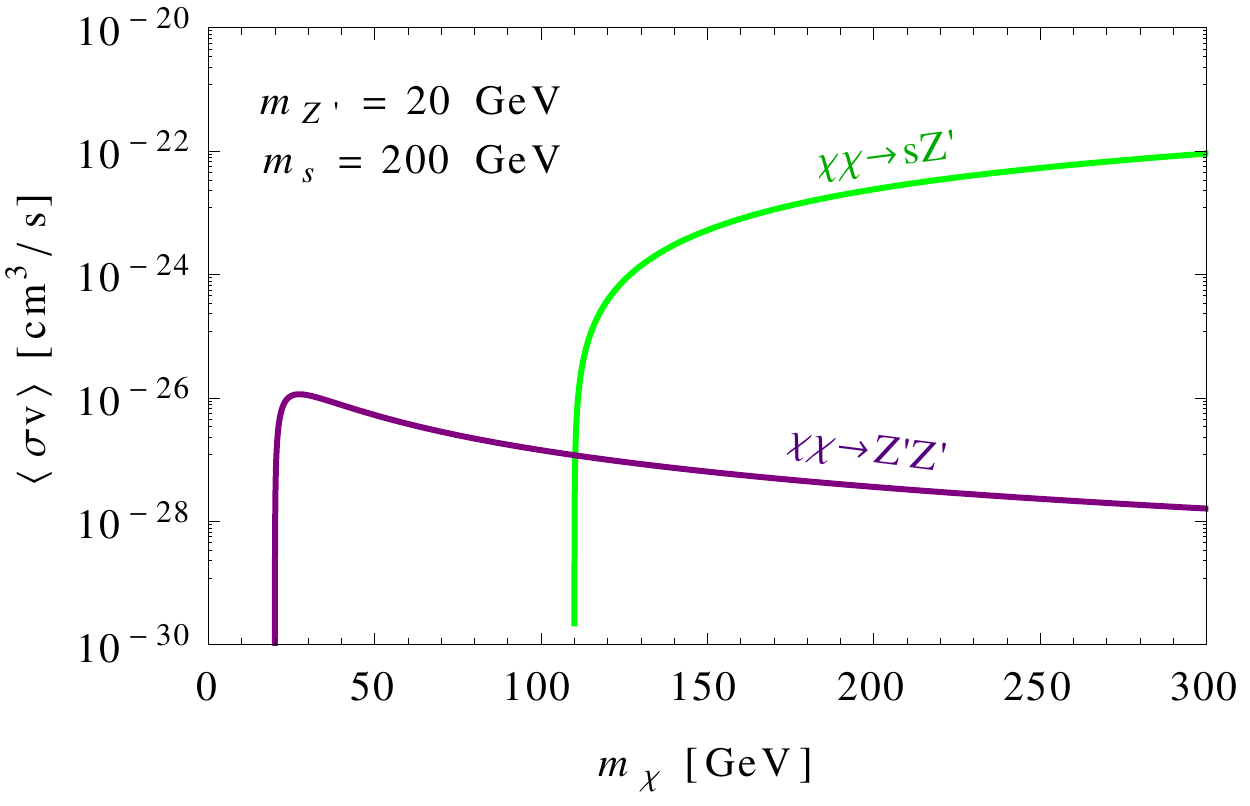}}
    \end{center}
    \caption{Relative cross section sizes for the two dominant
      $s$-wave diagrams, $\chi {\chi}\rightarrow s Z'$ (green)
      and $\chi {\chi}\rightarrow Z' Z'$ (purple), for some
      example parameter choices for the dark Higgs mass $m_s$ and the
      $Z'$ mass $m_{Z'}$, as labelled on each plot. For all plots the
      gauge coupling is set to $g_\chi=0.1$, but as all cross sections
      are directly proportional to $g_\chi^4$ they can easily be
      scaled by adjusting this parameter.  Note the lower two plots
      have a different $m_\chi$ range to the upper plots, so that the
      $y_\chi$ coupling is restricted to $\mathcal{O}$(1) values.}
   \label{fig:xsecs}
\end{figure}

As this new $s$-wave annihilation process is a consequence of
enforcing perturbative unitarity at high energies, its presence is
inevitable for axial-vector $Z'$-DM couplings. This means that the
limits on indirect detection signals using the $Z'Z'$ process alone
will lead to inaccurate conclusions. This can be seen in
Fig.~(\ref{fig:xsecs}), where we plot the annihilation cross sections
to both the $Z'Z'$ and $sZ'$ final states.  If the $s$ is lighter than
the $Z'$, there are values of DM mass $m_s + m_{Z'} < 2m_\chi <
2m_{Z'}$ where $sZ'$ is the only kinematically accessible final state.
If we were to only consider the $Z'Z'$ process, it would not be
possible to produce a limit for these low DM masses (where, in fact,
the indirect detection limits are the strongest).  When both $sZ'$ and
$Z'Z'$ are kinematically accessible, $sZ'$ becomes the dominant
process.  In the limit $m_\chi^2 \gg m_{Z'}^2,m_s^2$, the cross section
to $sZ'$ is enhanced relative to that for $Z'Z'$ by a factor of
$(m_\chi/m_{Z'})^4$, arising due to the longitudinal $Z'$
polarization.  It is important to note, however, that the DM mass and $Z'$ mass
are related via the dark Higgs vev, and thus satisfy
Eq.~(\ref{eq:yukawa}).  As a result, it is not possible to make the DM
mass arbitrarily large while retaining a perturbative value for the
Yukawa coupling $y_\chi$. For the mass ranges plotted in
Fig.~(\ref{fig:xsecs}), we have ensured that all parameters take
reasonable values.

The $s$-wave annihilations to $sZ'$ and $Z'Z'$ are by far the dominant
processes for indirect detection, for which the total annihilation
cross section is obtained by summing the contributions from these
channels.  In setting indirect detection limits, the energy spectra
should be computed by properly combining the spectra arising from the
$sZ'$ and $Z'Z'$ final states.  These $s$-wave processes will also be
the most important for the determination of relic density at
freezeout.  However, $p$-wave processes will also play a role at
freezeout, where the DM relative velocity is much larger than in the
universe today.  Note that as the cross sections in
Fig.~(\ref{fig:xsecs}) each scale as $g_\chi^4$, the correct thermal
relic density can easily be obtained simply by adjusting the value of
the dark gauge coupling.

 \subsection{Decay Widths of the Dark Higgs and $Z'$}
 
To compare our annihilation processes to indirect detection signals,
it is necessary to first multiply the thermal averaged cross sections
for our on-shell processes by relevant branching fractions. The $Z'$
decays to SM states via the hypercharge portal, and so couplings to all
fermion flavors must be considered.  The partial decay width of the
$Z'$ into SM fermions is given by
  \begin{equation}
   \Gamma(Z'\rightarrow f\bar{f})=\frac{m_{Z'}N_c}{12\pi}\sqrt{1-\frac{4m_f^2}{m_{Z'}^2}}\left[g_{f,V}^2\left(1+\frac{2m_f^2}{m_{Z'}^2}\right)+g_{f,A}^2\left(1-\frac{4m_f^2}{m_{Z'}^2}\right)\right],
   \label{eq:zppartial}
 \end{equation}
where $N_c$ is a color factor, relevant for hadronic decays. The $g_{f,V}$ (vector) and $g_{f,A}$ (axial-vector) coupling structures of the $Z'$ to the SM
are inherited from the kinetic mixing with the SM. The total decay width for the $Z'$ is simply the sum of all the fermionic decays,
  \begin{equation}
   \Gamma_Z'=\sum_f \Gamma(Z'\rightarrow f\bar{f}).
   \label{eq:zpwidth}
 \end{equation}
 The dark Higgs decays to the SM due to mass mixing with the SM
 Higgs. As it couples to fermions through their mass, the decay will
 be predominantly to $b$ quarks in the mass ranges we are considering,
 however we include all SM final states for accuracy.  The dark Higgs
 is also permitted to decay to pairs of $Z'$, although for simplicity
 we will choose parameters where this decay is not kinematically
 permitted. As loop decays and higher order corrections can be
 relevant for the dark Higgs decays, to ensure an accurate calculation
 of the branching fractions, we use the {\sc Fortran} package {\sc
   HDecay} \cite{Djouadi:1997yw}, which takes these effects into
 account.

\section{$\gamma$-ray Energy Spectra}
\label{sec:spectra}

The gamma ray flux $\Phi$ from photons with energy $E_\gamma$ resulting from dark matter annihilation into a fermion species $f$ is
  \begin{equation}
\frac{d^2\Phi}{d\Omega dE_\gamma}=\frac{\langle\sigma v\rangle}{8\pi m_\chi^2} \left( \sum_f \frac{dN}{dE_\gamma} Br_f \right) J(\phi,\gamma),
 \end{equation}
where $Br_f$ is the branching fraction to the particular fermion species. For the $Z'$ we take this as the ratio of Eq. (\ref{eq:zppartial}) and Eq. (\ref{eq:zpwidth}).
For the dark Higgs, we generate values using {\sc HDecay} \cite{Djouadi:1997yw}. The $J$ factor is the integral over the line of sight of the DM density $\rho(r)$ squared, 
at a distance $r$ from the center of the galaxy \cite{Cirelli:2010xx},
  \begin{equation}
J(\phi,\gamma)=\int\rho^2(r)dl,
 \end{equation}
 where we take the DM density to be modelled by the Navarro-Frenk-White (NFW) profile.

\begin{figure}[b]
\centering
\subfigure{\includegraphics[width=0.499\columnwidth]{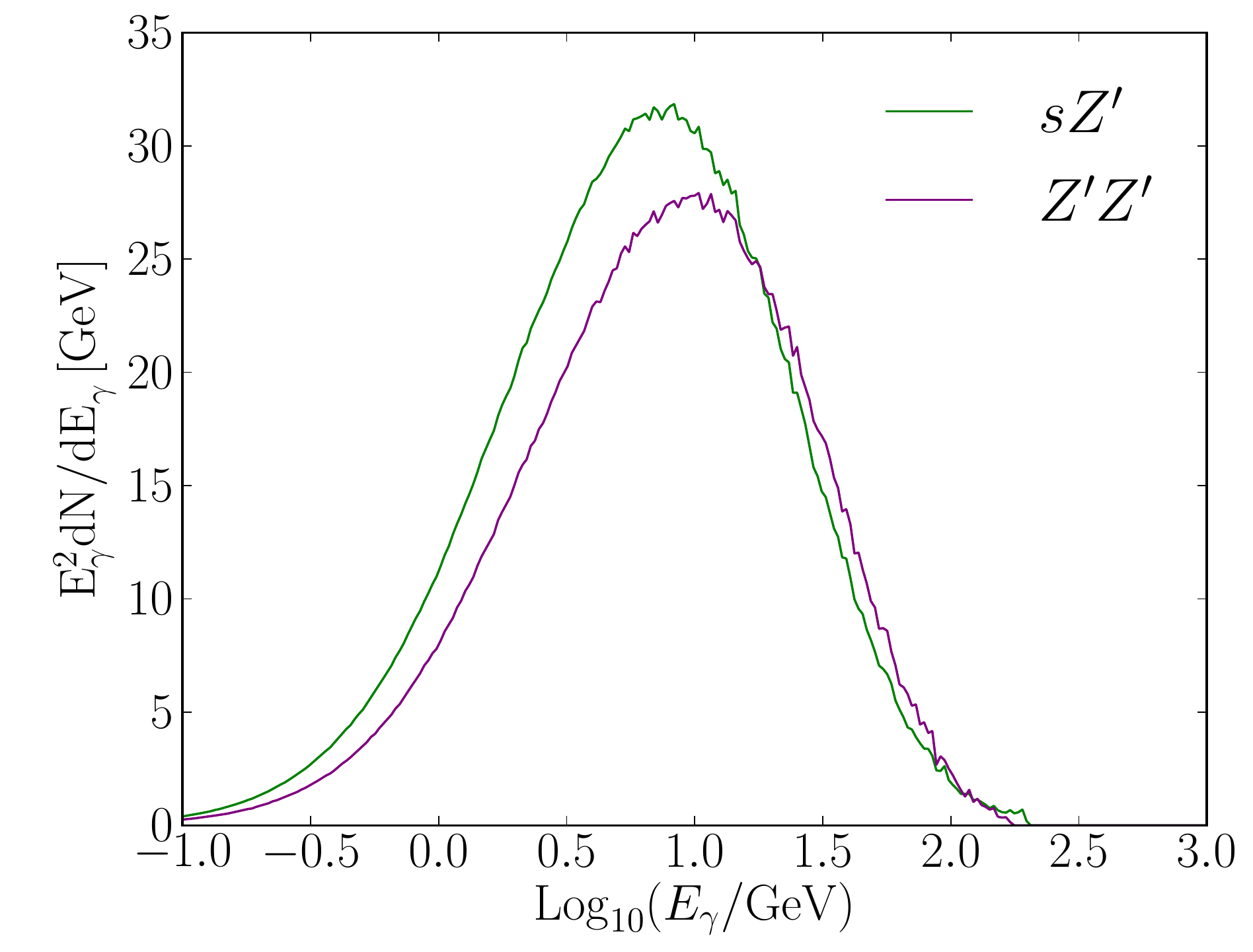}}
\subfigure{\includegraphics[width=0.485\columnwidth]{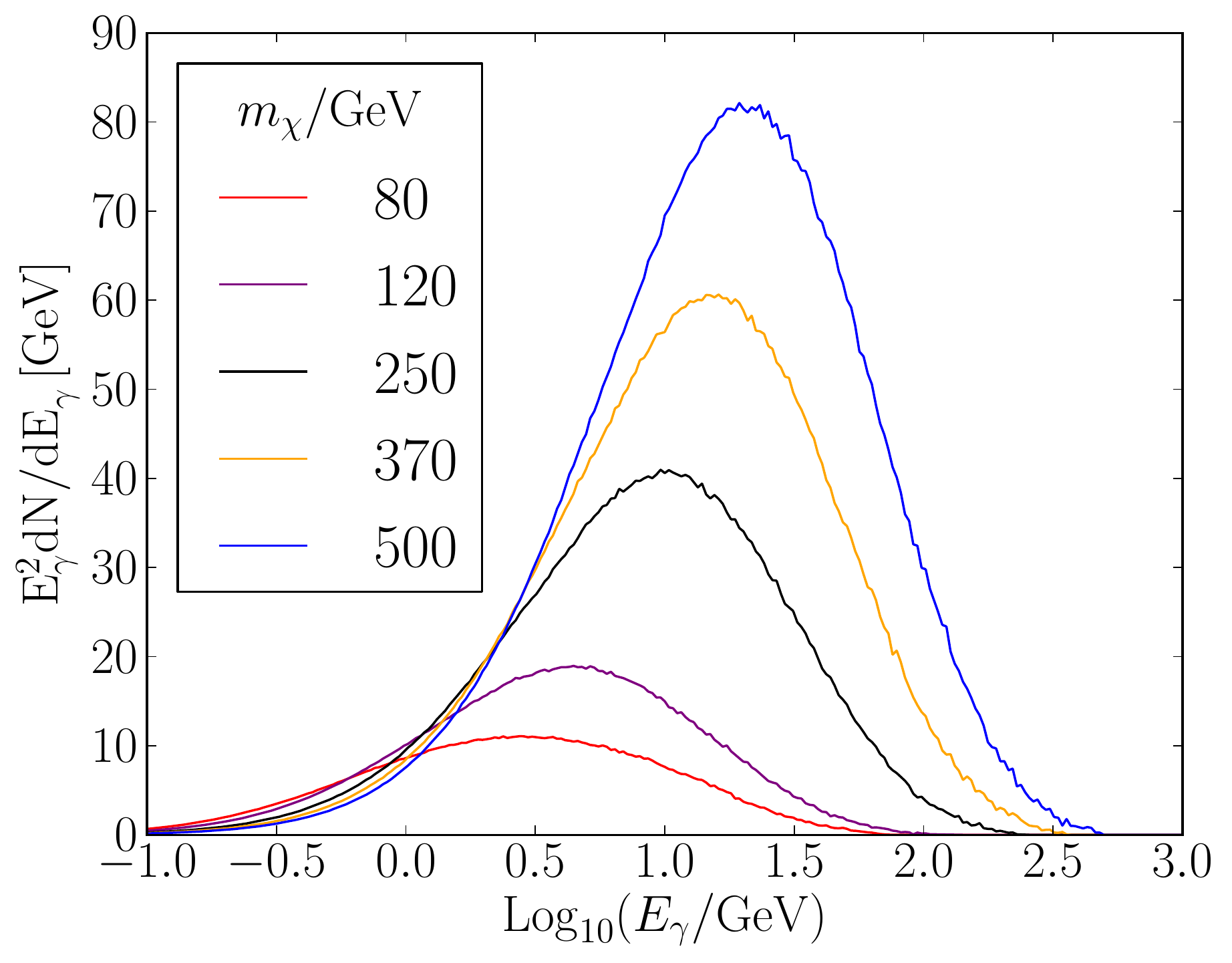}}
\caption{Left: Comparison of gamma ray spectra for DM annihilation into $sZ'$ vs. $Z'Z'$ for example parameters $m_s=100$ GeV, $m_{Z'}=60$ GeV and $m_\chi=200$ GeV. Right: Gamma ray spectra for DM annihilation to $sZ'$ with $m_s=30$ GeV and $m_{Z'}=120$ GeV,
for various DM masses. These plots include decays to all SM final states.}
\label{fig:gammaspec1}
\end{figure}

To obtain our $\gamma$-ray spectra, we simulate the annihilation cascade for a given DM mass with an effective resonance in {\sc Pythia} \cite{Sjostrand:2014zea}. 
In our setup, it is possible to have two different on-shell states which decay to SM fermions: the $Z'$ and the dark Higgs. To model for our
different states, we produce one diagram with two $Z'$ and one with two dark Higgs, both with effective resonances in their center of mass frames. 
We then average these to produce the effective spectra for a given DM mass. Specifically,
the effective resonances for different $Z'$ and dark Higgs $s$ masses are respectively given by \cite{Agashe:2014kda}
  \begin{equation}
E^{Z'}_{CoM}=\frac{s+m_{Z'}^2-m_s^2}{2\sqrt{s}}, \hspace{3mm} E^s_{CoM}=\frac{s+m_{s}^2-m_{Z'}^2}{2\sqrt{s}}.
 \end{equation}
Example gamma ray spectra including all possible fermionic SM final
states are shown in Fig.~(\ref{fig:gammaspec1}), as well as a
comparison of the $sZ'$ and $Z'Z'$ spectra for example parameters.

\section{Annihilation Limits from Dwarf Spheriodal Galaxies and AMS-02}
\label{sec:dwarfs}

Currently, two of the strongest constraints on dark matter
annihilation processes come from AMS-02, for low DM masses and
electron-positron final states, and from Fermi-LAT limits placed on
signals from dwarf spheriodal satellite galaxies of the Milky Way
\cite{Ackermann:2015zua}.  Dwarf spheriodal galaxies (dSphs) are
particularly useful in constraining dark matter models, as according
to kinematic data they are one of the most dark matter dense objects
in the sky, and have relatively low backgrounds.  However, the limits
published by Fermi-LAT assume a 100$\%$ branching fraction to a
particular SM final state, and within our kinetically mixed $Z'$ model
this will not be true due to the flavor universal nature of the
mixing. It is also not trivial to simply scale the dSphs limits with
our branching fractions, as not only are the kinematics are different, but
as there can be cross-polution of photon contributions from different
final states. Furthermore, our new process
$\chi {\chi}\rightarrow s Z'$ has two different final state
particles with different masses, and the resulting spectra will depend
on the specific masses of these particles.  Therefore it is necessary
to recast the limits for this specific setup, comparing to the dSphs
likelihood functions released by Fermi-LAT.

\begin{figure}[h]
\centering
\subfigure{\includegraphics[width=0.48\columnwidth]{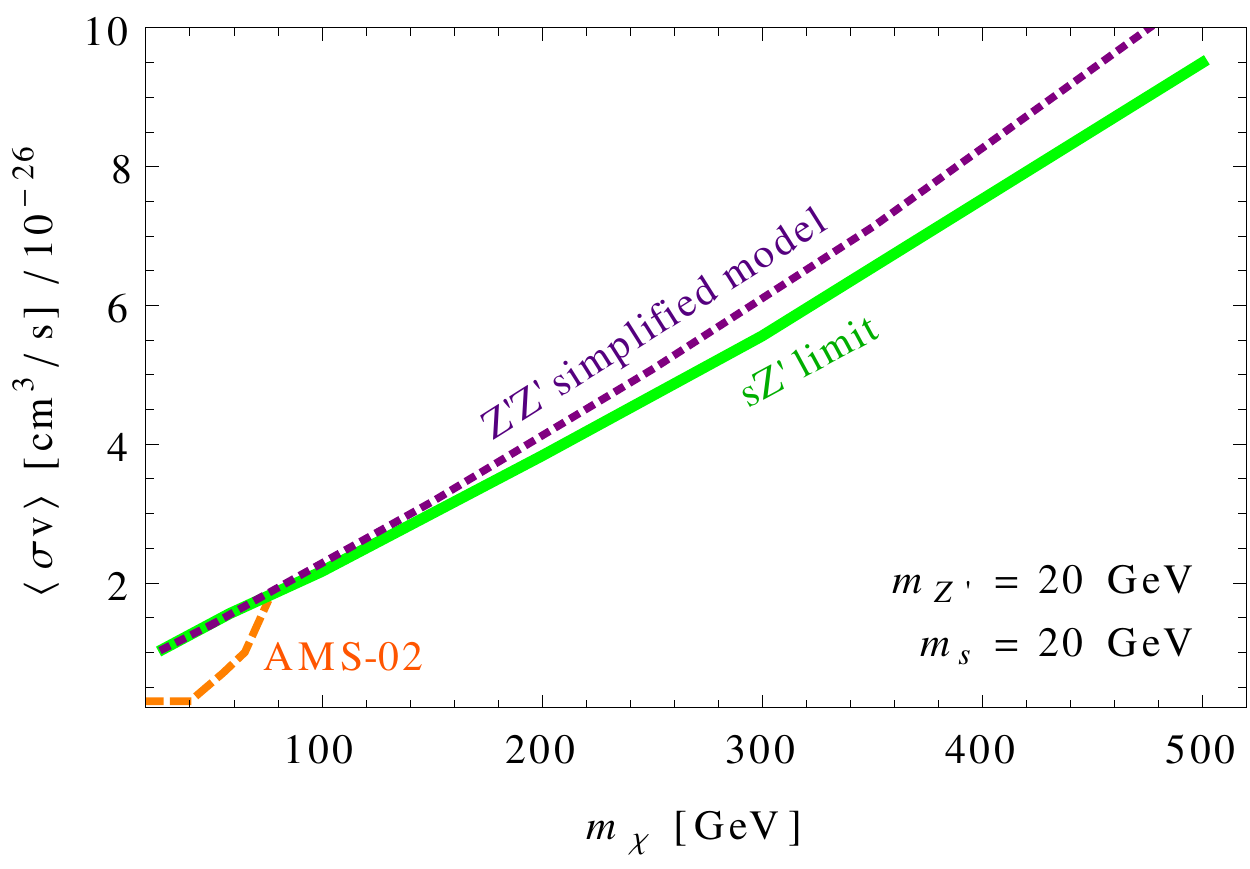}}
\subfigure{\includegraphics[width=0.48\columnwidth]{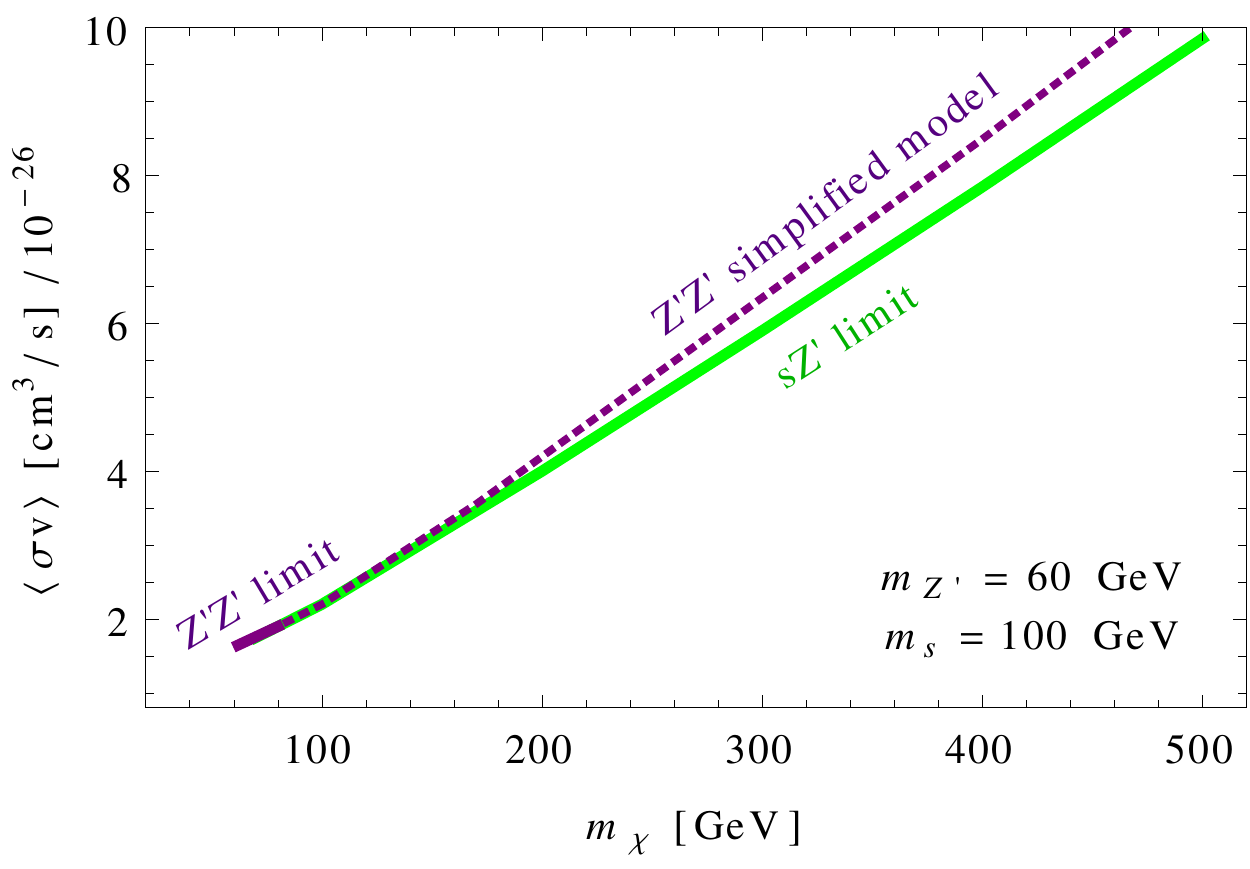}}\\
\subfigure{\includegraphics[width=0.48\columnwidth]{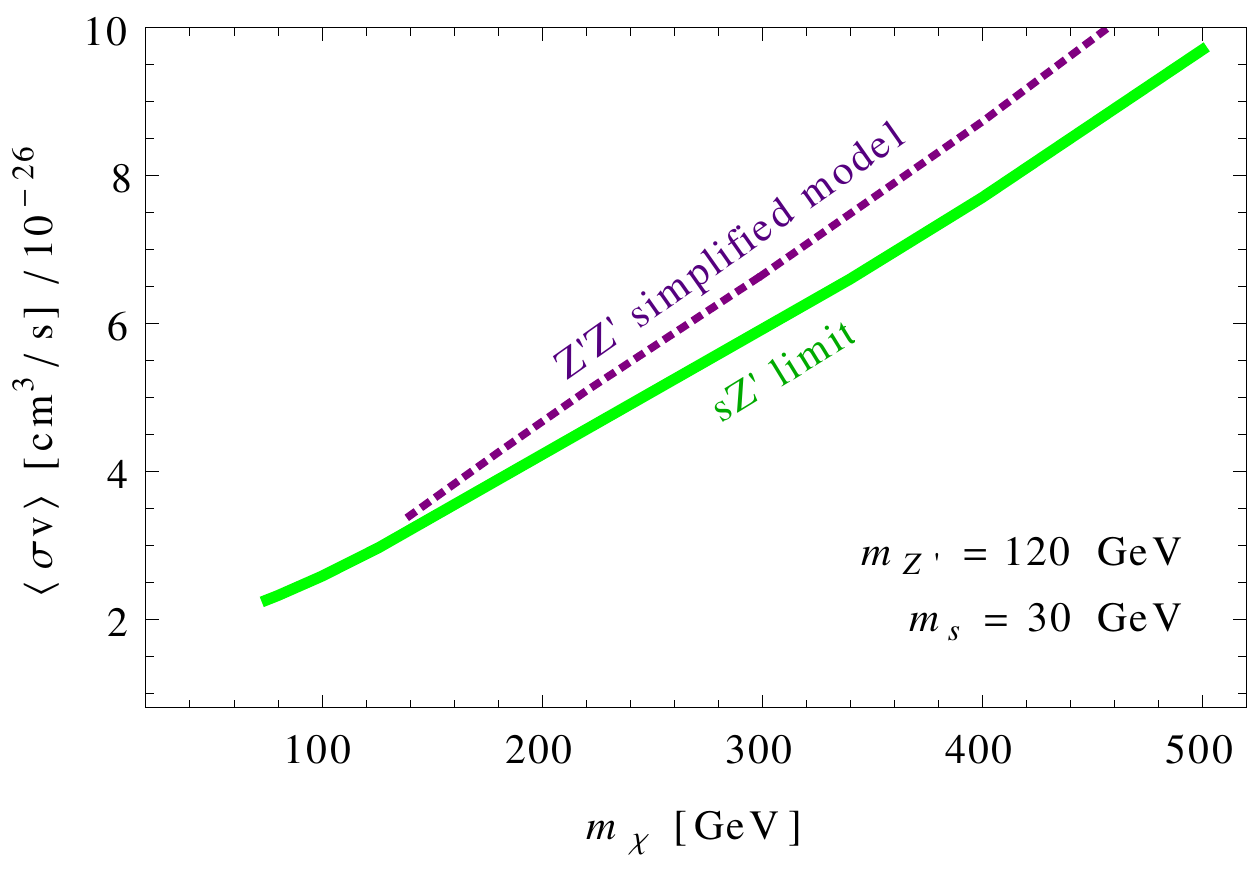}}
\subfigure{\includegraphics[width=0.48\columnwidth]{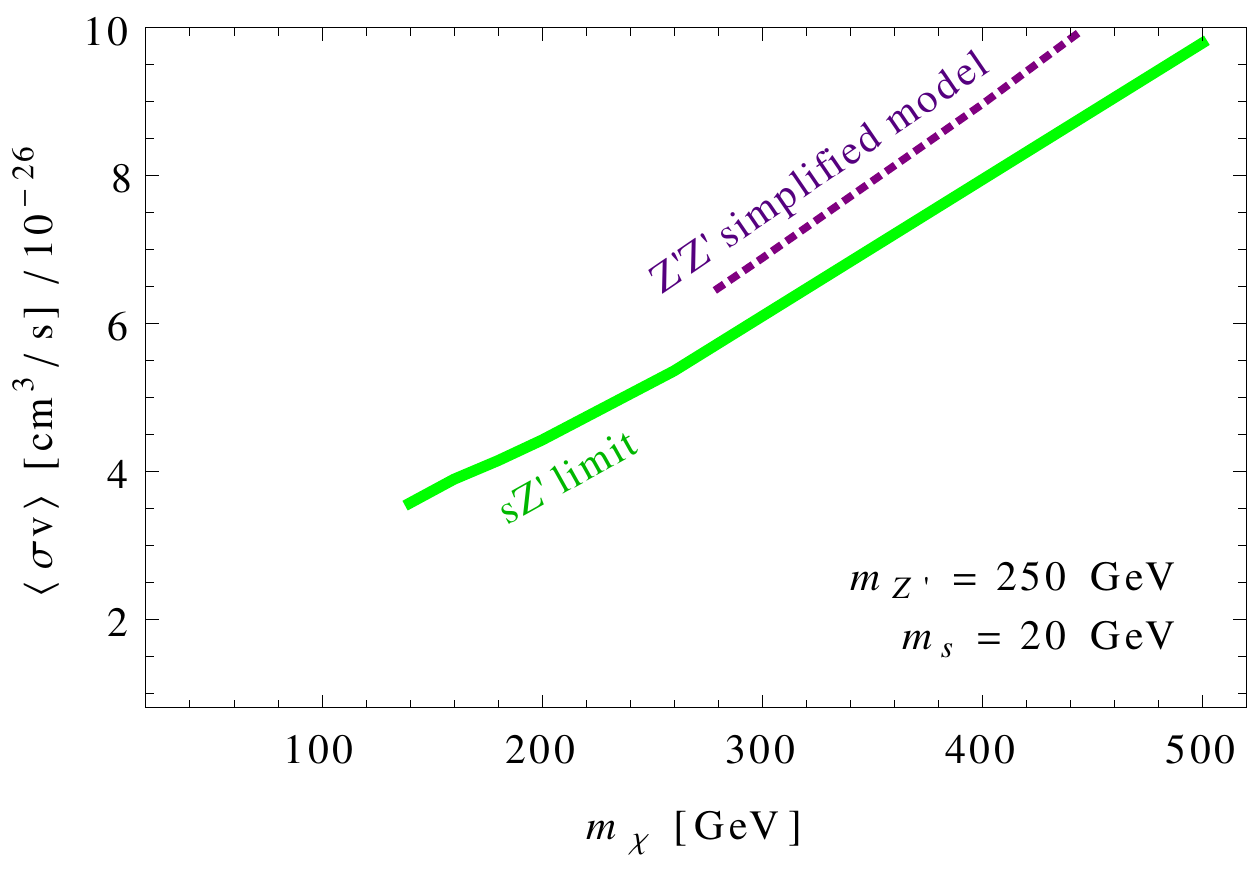}}
\caption{95$\%$ confidence limits (C.L.) on the annihilation cross
  section from Fermi data on 15 dwarf spheroidal galaxies.  All solid
  lines are limits on our model: the purple line is the cross section
  limit arising from the $Z'Z'$ process is alone; the green line is
  the cross section limit for the $sZ'$ process alone. The purple dotted line is the $Z'Z'$ limit
  alone as per the simplified model with no dark Higgs. The
  approximate limit from AMS-02 is shown in orange. Masses are as
  labelled in each plot.}
\label{fig:fermix1}
\end{figure}

To find the limit on the cross section from dSphs, we use our spectra
generated with {\sc Pythia} \cite{Sjostrand:2014zea}, as described
in the previous section. We then use the maximal likelihood method to
compare our spectra against those for the dSphs provided by
Fermi-LAT, with the $J$ factor taken to be a nuisance parameter. We
take spectra from 15 dSphs: Bootes I, Canes Venatici II, Carina, Coma
Berenices, Draco, Fornax, Hercules, Leo II, Leo IV, Sculptor, Segue 1,
Sextans, Ursa Major II, Ursa Minor, and Willman 1.  The 95$\%$
C.L. limits on the annihilation cross section from dSphs for both
$Z'Z'$ and $sZ'$ spectra are shown for some example parameters in
Fig.~(\ref{fig:fermix1}).

In Fig.~(\ref{fig:fermix1}), limits are set on the individual $sZ'$
and $Z'Z'$ cross sections.  In general, the limits arising from the
spectral shape of the DM annihilation to $sZ'$ is slightly more
constraining than that from $Z'Z'$.  This is likely due to the peak of
the gamma ray spectra produced by the scalar being higher than that
produced by the $Z'$. Which limit is relevant depends on which of the
final states is kinematically accessible.  When $sZ'$ is accessible it
greatly dominates, and hence the cross section limit is given by the
solid green $sZ'$ line.  If $Z'Z'$ but not $sZ'$ is accessible then the
solid purple $Z'Z'$ line shows the relevant limit.  The purple dotted
line corresponds to the limit on annihilation to $Z'Z'$ alone, as
would occur in a simplified model with only a $Z'$ mediator and no
dark Higgs. This allows for a comparision of the simplified model with our
scenario.

To find the limit from AMS-02, it is sufficient to only consider
electron-positron final states, as these provide the strongest
limits. As the dark Higgs couples to particles through their mass,
there will be negligible production of electron final states via decay
of the $s$.  This means that the $Z'$ decays will provide effectively
all the electron-positron signal. In the low DM mass range, where
AMS-02 is most constraining, the limit on the cross section is
approximately flat for cascade decays to two identical final state
particles \cite{Elor:2015bho}. Therefore, we scale the cross section
limit on electron final states by the branching fraction of $Z'$ to
electron-positron pairs. This is stronger than the dSphs limit only
for low DM masses (and hence low $s$ and $Z'$ masses).  As a result,
AMS-02 limits are relevant only for low mass parameters, and shown
on only one of the plots of Fig.~(\ref{fig:fermix1}) for which the
$Z'$ and $s$ masses are both small.

\section{Other Model Constraints}
\label{sec:other}

The indirect detection constraints are determined purely by the
couplings of the mediators to DM, controlled by $g_\chi$, and the mass
parameters $m_\chi$, $m_{Z'}$ and $m_s$.  The exact size of the small
couplings of the mediators to SM fermions, controlled by the mixing
parameters $\epsilon$ and $\lambda_{hs}$, does not affect the indirect
detection signals, as the mediators have long astrophysical time
scales over which to eventually decay.  However, other experimental
probes, such as direct detection and collider experiments, are
directly sensitive to the size of the small dark-SM couplings.

\subsection{Collider and Direct Detection Constraints}

As the couplings between the dark and visible sectors are taken to be
very small, it is possible to completely escape the strong WIMP DM
constraints from the LHC and direct detection.  This provides a
compelling scenario which is consistent with the null results of these
experiements to date, while still allowing a large indirect detecion
signal.

\subsection{BBN and CMB Constraints}

A lower limit on the size of the couplings between the sectors comes
from Big Bang Nucleosynthesis (BBN), which requires that the mediators
have a lifetime of $\tau<1$s \cite{Chen:2009ab}. This leaves
a large range of values (several orders of magnitude) for the
kinematic mixing parameter $\epsilon$ and Higgs portal parameter
$\lambda_{hs}$.  In addition, CMB measurements can also provide constraints
on the annihilation cross sections, however they are weaker than those
arising from AMS-02 and dSphs \cite{Elor:2015bho}.

\subsection{Unitarity}
As discussed above, the dark Higgs is included not only to provide
a mass generation mechanism for the dark sector, but to ensure
perturbative unitarity is not violated at high energies.  In the
absence of the scalar, unitarity violation would arise at high energy
due to the longitudinal mode of the $Z'$ gauge boson in processes such
as $\chi {\chi}\rightarrow Z'Z'$.

In the indirect detection context, where it is appropriate to take the
zero velocity limit, it turns out that the cross section for
$\chi {\chi}\rightarrow Z'Z'$ receives no contribution from
the scalar exchange diagram of Fig.~(\ref{fig:zpzp}).  However, at high
energies where the $v=0$ threshold approximation is no longer valid
(including at freezeout) the scalar diagram cannot be
neglected~\cite{Cline:2014dwa}.  Regardless, the scalar is mandatory
in any model in which the $Z'$ has axial-vector couplings to fermions,
in order to properly respect gauge invariance and perturbative
unitarity~\cite{Kahlhoefer:2015bea}.

\section{Summary}
\label{sec:summary}

We have considered a self-consistent dark sector containing a Majorana
fermion DM candidate, $\chi$, a dark gauge boson, $Z'$, and a dark
Higgs, $s$, which transform under a dark $U(1)_{\chi}$ gauge symmetry.
This is the minimal consistent model in which a Majorana DM candidate
couples to a spin-1 mediator.  In this scenario, the coupling of the
DM to the $Z'$ must be of axial-vector form, as vector couplings of
Majorana fermions vanish.  The dark Higgs field provides a mass
generation mechanism for both the $Z'$ gauge boson and the DM $\chi$,
and is required in order for the model to properly respect gauge
invariance and perturbative unitarity.

We have investigated the indirect detection phenomenology of this
model, focusing on the processes where the DM annihilates to on-shell
dark sector mediators.  We found that the presence of a spin-0 and
spin-1 mediator in the same model opens up an important new $s$-wave
annihilation channel, $\chi \chi\rightarrow sZ'$, which can dominate
over the well-studied process $\chi \chi\rightarrow Z'Z'$.  This is to
be contrasted to the situtation in simplified models that contain a
single mediator: there is no $s$-wave annihilation process to scalar
mediators; $s$-wave annihilation to pseudoscalar mediators is
suppressed by 3-body phase space; the process $\chi \chi\rightarrow
Z'Z'$ is the only $s$-wave annihilation to vector or axial-vector
mediators (which, in the case of an axial mediator, violates unitarity
at high energy).  The inclusion of the scalar and vector mediator in
the same model allows sizable production of the scalar mediator via
$s$-wave annihilation, which was previously not thought possible, and
provides a very plausible way to discover the dark Higgs.  This
important phenomenology is missed in the single-mediator simplified
model approach.

We have calculated indirect detection limits on the $sZ'$ and $Z'Z'$
annihilation processes, using Fermi-LAT gamma ray data for dwarf
spheriodal galaxies.  The gamma ray energy spectra resulting from the
two annihilation modes are similar.  Depending on the masses of the
dark sector particles, there are regions of parameter space where only
one of the $sZ'$ and $Z'Z'$ final states are kinematically accessible.
As such, the new process allows a broader range of DM masses to be
probed via indirect detection.  In the limit that $m_\chi^2 \gg
m_{Z'}^2,m_s^2$, where both processes are kinematically allowed, the
cross sections to $sZ'$ is much greater than that to $Z'Z'$.
Neglecting the $sZ'$ process, as done in the simplified model setup,
would lead to highly inaccurate constraints on the model parameters.

An important observation is that the mass and coupling parameters in
the dark sector may be intrinsically related to each other.  In our
case, the various parameters are related via the gauge coupling
constant and the dark Higgs vev, such that we do not have the freedom
to vary all parameters independently.  The absence of this feature is
one of the shortcomings of the (albeit very useful) simplified model
approach.  In general, renormalizable models in which gauge invariance
is enforced will be a superior approach.  Not only are unitarity
problems avoided, but the phenomenology is potentially richer.

\section{Acknowledgements}

This work was supported in part by the Australian Research
Council. Feynman diagrams are drawn using {\sc TikZ-Feynman}
\cite{Ellis:2016jkw}.  We thank the authors of
Ref.~\cite{Duerr:2016tmh} for pointing out an error in an earlier
version of our cross section.

\bibliographystyle{JHEP}
\bibliography{darkhiggs_30jun.bib}

\end{document}